# Synchronization of two anharmonic nanomechanical oscillators

M. H. Matheny, M. Grau, L. G. Villanueva, R. B. Karabalin, M. C. Cross, M. L. Roukes

*Kavli Nanoscience Institute and Departments of Physics, Applied Physics, and Bioengineering,*

*California Institute of Technology, Pasadena, California 91125*

We investigate the synchronization of oscillators based on anharmonic nanoelectromechanical resonators. Our experimental implementation allows unprecedented observation and control of parameters governing the dynamics of synchronization. We find close quantitative agreement between experimental data and theory describing reactively coupled Duffing resonators with fully saturated feedback gain. In the synchronized state we demonstrate a significant reduction in the phase noise of the oscillators, which is key for sensor and clock applications. Our work establishes that oscillator networks constructed from nanomechanical resonators form an ideal laboratory to study synchronization – given their high-quality factors, small footprint, and ease of co-integration with modern electronic signal processing technologies.

Synchronization is a ubiquitous phenomenon both in the physical and biological sciences. It has been observed to occur over a wide range of scales - from the ecological[1], with oscillation periods of years, to the microscale[2], with oscillation periods of milliseconds. Although synchronization has been extensively studied theoretically[3-5], relatively few experimental systems have been realized that provide detailed insight into the underlying dynamics. Here we show that oscillators based on nanoelectromechanical systems (NEMS) can readily enable the resolution of such details, while providing many unique advantages for experimental studies of nonlinear dynamics[6-8].

Nanomechanical oscillators also have been exploited for a variety of applications. In particular, nanoscale mechanics exhibits enhanced nonlinearity[9] and tunability[10], which has been used to suppress feedback noise[11, 12] and create new types of electromechanical oscillators[13, 14]. These oscillators may find application as mass[15], gas[16, 17], or force[18] sensors, without the need of an external frequency source. In addition to their extreme sensitivity, they dissipate very little power due to their high quality factors, reducing the sustaining power needed for sensor arrays.

Although NEMS arrays can provide exceptional performance as frequency-shift sensors or frequency sources, their implementation can be challenging. For example, statistical deviations in batch fabrication inevitably lead to undesirable array dispersion[16]. If a sensor array has appreciable frequency dispersion, global sensor responsivity gets reduced due to an overall increase in signal phase noise. However, upon synchronization, dispersive elements lock to a single frequency. If the oscillators are not only frequency locked, but phase locked, the phase noise of this array may be reduced[3]. Attainment of this can mitigate the deleterious effects from an array's frequency dispersion. Since NEMS have numerous applications, and are useful in studying nonlinear dynamics, we set an important milestone by demonstrating synchronization in nanomechanical systems.

There are previous reports[19, 20] of synchronization in micro- or nanomechanical systems. However, these do not, in fact, demonstrate the phenomenon as conventionally defined[3] – that is, the phase locking of weakly coupled self-sustained oscillators. Shim, *et al.* reported synchronization of the driven excitations in coupled resonators, not self-sustained oscillators. Zhang, *et al.* reported self-sustained oscillations excited by radiation pressure in optomechanical resonators, coupled through the evanescent optical field. However, the model and data presented in Zhang, *et al.* reflect strong coupling[21], with the energy to excite the oscillations equal to the energy to couple the devices. This strong coupling inevitably leads to confusion between synchronization of individual oscillators and the excitation of a single coupled mode.

Our experiment is designed to unambiguously demonstrate canonically-defined synchronization with a pair of weakly coupled oscillators. This is accomplished by employing an additional feedback loop, separate from the feedback loop necessary to sustain oscillations, to couple the resonant devices. This coupling feedback can be modified via electronic attenuation and phase shifting, allowing for full control of all relevant parameters. Importantly, it can set to a value where the coupling is a weak perturbation on the individual oscillator dynamics. We also have precise control over the other system parameters, detuning and frequency pulling, described below. Since all of these parameters are carefully calibrated, we can make quantifiable comparisons with theoretical models, yielding an ideal platform to elucidate synchronization phenomena. Our implementation is readily scalable to thousands of devices through standard methods of large-scale integration. In order to show the applicability of synchronized NEMS, we measure the phase noise of the oscillators, and demonstrate the reduction in phase noise theoretically expected from noise averaging.

Synchronized networks fall into two separate classes based on the type of interactions between elements[3, 22]. These interactions, the oscillator coupling, can be either dissipative or reactive (or a combination thereof). To date, most studies of synchronization have focused on dissipative coupling in a reduced single-variable description, i.e., the Kuramoto model[23]. In many systems with dissipative coupling, relative amplitude differences do not affect the synchronization, and are ignored. However, by contrast, many natural synchronized systems display reactive coupling[24], where the differences in individual oscillator amplitudes enable their synchronization. In the experiments reported here, we focus on reactive coupling and measure oscillator frequencies and amplitudes. In manufactured systems, it may be advantageous to avoid dissipative coupling since this will often introduce additional noise into the system (via the fluctuation-dissipation theorem) and therefore degrade the frequency precision of the synchronized state. Reactive coupling has been demonstrated in NEMS,[25, 26] where it can be created straightforwardly through electrostatic or mechanical means. Previous theoretical work shows that large arrays can synchronize through an interaction of anharmonicity inherent in NEMS devices with this reactive coupling[24]. This present work is a first milestone in the experimental investigation of synchronization where both large scale behavior and individual elements can be simultaneously controlled and observed in detail.

We describe our system with a set of equations similar to the model theoretically examined by Aronson *et al.*[27], except that here our system amplitude is not constrained by nonlinear dissipation, but rather by amplifier saturation. We scale the amplitude in our equations by the level of saturation, and examine the system dynamics in "slow" time, $T \sim Q * t * \omega_o$, where $Q$ is the quality factor of the driven response of the resonators and $\omega_o$ the linear resonance frequency of the NEMS device when under driven excitation, and t is the real time in seconds. In the slow time dynamics, feedback loop time delays are represented by a phase shift. The resulting equations for the amplitudes $a_{1,2}$ for each oscillator and phase difference $\varphi$ between them are[21]

$$a_1' \equiv \frac{da_1}{dT} = -\frac{a_1}{2} + \frac{1}{2} - \frac{\beta}{2} a_2 \sin \varphi, \qquad (1)$$

$$a_2' \equiv \frac{da_2}{dT} = -\frac{a_2}{2} + \frac{1}{2} + \frac{\beta}{2} a_1 \sin \varphi, \qquad (2)$$

$$\varphi' \equiv \frac{d\varphi}{dT} = \Delta\omega - (a_1^2 - a_2^2)\left(\alpha - \frac{\beta}{2a_1 a_2} \cos\varphi\right). \qquad (3)$$

Here $\Delta\omega$ is the difference between the resonant frequencies of the devices, $\alpha$ is the measure of the amount of frequency pulling (which is the increase in frequency proportional to the square of the amplitude), and $\beta$ is the coupling strength. The parameters $\Delta\omega, \alpha,$ and $\beta$, which we call the *synchronization parameters*, set the dynamics of the system: the stable fixed points of equations 1-3, for example, yield synchronized states. These parameters are expressed in units of the device's resonance line width, $\omega_0/Q$. For example, $\Delta\omega = 1$ corresponds to a resonator frequency difference of 1 line width. Note when the coupling term is not present ($\beta = 0$), we obtain a fixed point for equations 1-3 such that $a_1 = a_2 = 1$ and $\varphi' = \Delta\omega$. Therefore, by measurement of the uncoupled oscillator amplitudes and frequency differences we can calibrate the frequency pulling $\alpha$, and detuning $\Delta\omega$.

In order to construct an experiment with independent control of the synchronization parameters we use the setup shown in Figure 1. The NEMS devices are two piezoelectrically actuated, piezoresistively detected,[14] doubly-clamped beams $10\ \mu m$ long, $210\ nm$ thick, and $400 nm$ wide. The signal from each beam is split into two different feedback loops. One feedback loop sets the level of oscillations (the *oscillator loop*), and the other loop sets the coupling (the *coupling loop*). In the oscillator loop, the signal is strongly amplified (gain stage, $g$) into a diode limiter (saturation stage, $s$) in order to ensure the feedback signal to the beam is of constant magnitude[12]. Therefore, the feedback signal is a strongly nonlinear function of the device displacement. On the other hand, the coupling loop is kept linear; the feedback is directly proportional to the displacement over the full range of experimental values. For the oscillator loop, the signal is fed back in phase with the beam's velocity. For the coupling loop, this signal is fed into the beams in phase with the displacement. This causes the coupling branch to be reactive and the oscillator loop to be dissipative.

This system is designed to be integrable within CMOS technologies. The system consists of transistor amplifiers, saturation diodes, direction couplers (capacitors), and phase shifters. Here we use adjustable attenuators; however, these can be implemented with adjustable amplifiers. The phase shifters may be implemented with fixed RC filters; however, we note that if we measure the piezoelectric response in addition to the piezoresistive response, we are able to directly capture both the in-phase and out-of-phase response of the oscillators. This solution finds the most application when the feedback delay is small compared to the period of the signal with all the components integrated on chip.

It is important to note that the three parameter controls $(\Delta\omega, \alpha, \beta)$ are independent. This makes the experimental data easier to process, and helps clearly identify which modified parameter induces synchronization. More details can be found elsewhere[21].

We begin the discussion by looking at the small coupling limit, with the coupling less than a tenth of the resonator width, where experimental data can be compared to analytical predictions. In that case, the amplitudes of the two oscillators stay near unity (the fixed points of equations 1,2 give $a_{1,2} \approx 1 \mp \beta \sin\varphi$). In this limit, equation 3 assumes the form of the Adler equation[28]

$$\varphi' = \Delta\omega + 4\alpha\beta \sin\varphi. \qquad (4)$$

Note that even though this equation is of the same form as Adler's study of injection locking, our equation is describing the mutual synchronization of two oscillators[21]. When the oscillators are unsynchronized, its solution can expressed in terms of the oscillator frequency difference

$$\varphi' = \sqrt{\Delta\omega^2 - (4\alpha\beta)^2}. \qquad (5)$$

Equations 4 and 5 mimic an overdamped Josephson junction using the RCSJ model[29]. The oscillator phase difference $\varphi$ corresponds to the phase difference across the Josephson junction, the detuning $\Delta\omega$ to the injected DC current (normalized to the ratio of the junction's normal state resistance and a flux quantum), and the frequency pulling-coupling term $4\alpha\beta$ to the critical current (again normalized to the ratio of the junction's normal state resistance and a flux quantum). However, unlike the critical current which is fixed by junction geometry in the RCSJ model, here we can experimentally control both frequency pulling and coupling independently.

In Figure 2, we compare the analytical predictions of equations 5 and 6 with the experimental data for the amplitudes and frequency difference as the detuning is swept, with a fixed value of frequency pulling $\alpha = 1.25$. In Figure 2 $\varphi'$ is the oscillator frequency difference in units of the resonance width. In the synchronization regime, as the amplitudes stay near unity, a linear relationship between the oscillation amplitudes and the frequency difference is found from equation 4 for,

$$\frac{\Delta a_{1,2}}{a_{1,2}} = \pm \frac{\Delta \omega}{4\alpha}, \tag{6}$$

where 1,2 corresponds to +,-, respectively. The plots clearly show synchronization between the two coupled oscillators. The agreement between theoretical predictions, given by the Adler equation, and the experimental data is remarkable. Note that upon synchronization, the oscillator amplitudes change in order to adjust the oscillator frequencies. This shows that the frequency pulling is crucial to the synchronization for reactive coupling.

In addition to control of the detuning through a wide range of values (shown in Figure 2), we are able to modify both the frequency pulling and coupling, to study the parameter space for synchronization. Figure 3 shows the synchronization parameter space for three levels of fixed detuning ($\Delta \omega = 0.6, 1, 2$) as coupling and frequency pulling $\alpha$ are varied. The red border is the data with attractive coupling ($\beta < 0$ in equations 1-3) and green with repulsive coupling ($\beta > 0$ in equations 1-3). These lines represent the boundaries of the transition between synchronized and unsynchronized states when sweeping to higher values of coupling, i.e., from left to right in Figure 3. This transition is defined by a change to a measured oscillator frequency difference $\varphi' < 0.05$.

In general, analytical solutions to equations 1-3 cannot be found. Therefore, we perform two numerical studies and compare them to the experiment. We perform a linear stability analysis[30] (LSA) of equations 1-3 with the orange and purple dashed lines in Figure 3 showing the stability boundaries. We also perform a time domain simulation of equations 1-3, using initial conditions of amplitudes fixed at 1 and random phases. This time domain simulation gives a basin of attraction for stabilizing in either an unsynchronized or a synchronized state. For each value of the parameters plotted in Figure 3, we run 100 such simulations and assign a "synchronization value" between 0 for unsynchronized and 1 for synchronized. The average value of these 100 simulations is represented by a linear gradient between white and blue for 0 and 1, respectively..

We can see a clear distinction between the sets of experimental data corresponding to attractive and repulsive coupling (red and green lines). However, Equations 1-3 are completely symmetric upon exchange of $\beta \rightarrow -\beta$, since synchronization will occur for $\varphi \rightarrow \varphi + \pi$. The numerical time-domain simulation shows correspondence to the experimental data, indicating that the initial phase difference of the oscillators is not completely random. The lightest blue basins of attraction are those where the initial phase must be close to $\varphi_0 = \pi$ in order to synchronize. More initial phase conditions will synchronize as the basins transition from light to dark. When totally dark even initial phase conditions near $\varphi_0 = 0$ will synchronize. Thus the sign of coupling tends to bias the experiment to different initial phase conditions.

In the set of data with largest detuning, $\Delta \omega = 2$, the experiment shows somewhat larger departure from theoretical predictions. We observe that at large detunings, asymmetries in saturation level or discrepancies in quality factor between the two oscillators tend to create larger disagreement between theory and experiment. This is due to the large coupling necessary in order to synchronize the oscillators, which magnifies the nonlinear behavior (and thus asymmetry) of the system. However, the close agreement of Figures 2 and 3 show the generality and accuracy of our approach.

Finally, we explore the effect of synchronization on the phase noise. In Figure 4, the green and blue spheres are the phase noise at 1kHz offset from the carrier frequency (a key figure-of-merit for the frequency source community[31]) plotted as a function of coupling for oscillators 1 and 2, respectively[21]. The red diamonds show the oscillator frequency difference $\varphi'$ for comparison. As coupling is increased the phase noise at this offset initially increases (due to phase slipping between the oscillators) and then suddenly drops to 3dB below the uncoupled noise level. The plot of the oscillator frequency difference indicates the phase noise reduction occurs at the onset of synchronization. This corresponds to a phase noise reduction by factor of two, as predicted by theoretical estimates[3]. If acted upon by the same stimulus, this oscillator array would show an improvement in signal-to-noise. This would be useful for measurement of weak global quantities, such as gravitational fields[32] or gas environments[17].

Our demonstration of the synchronization of two reactively coupled anharmonic NEMS oscillators shows excellent agreement with analytical and numerical modeling. We track not only the frequency difference, but also the individual amplitudes, important for a full multivariable description of the synchronization. These results highlight the importance of the oscillator amplitudes in synchronization for reactive coupling. This work highlights the potential of nonlinear dynamics experiments at the intermediate scale of discretization: full control of individual elements and tracking of large arrays is now possible. All of the components in these experiments can be realized using CMOS technology, implying that very-large-scale networks can be built using the precise technology of present-day semiconductor nanoelectronics and electronically tested with co-integrated state-of-the-art signal processing capabilities. The flexibility of this system permits creation of dissipative or reactive coupling in arbitrarily complex or completely random networks. Our experimental demonstration of reduced phase noise in the synchronized state paves the way for detection of very weak phenomena using coupled nanoscale sensor arrays.

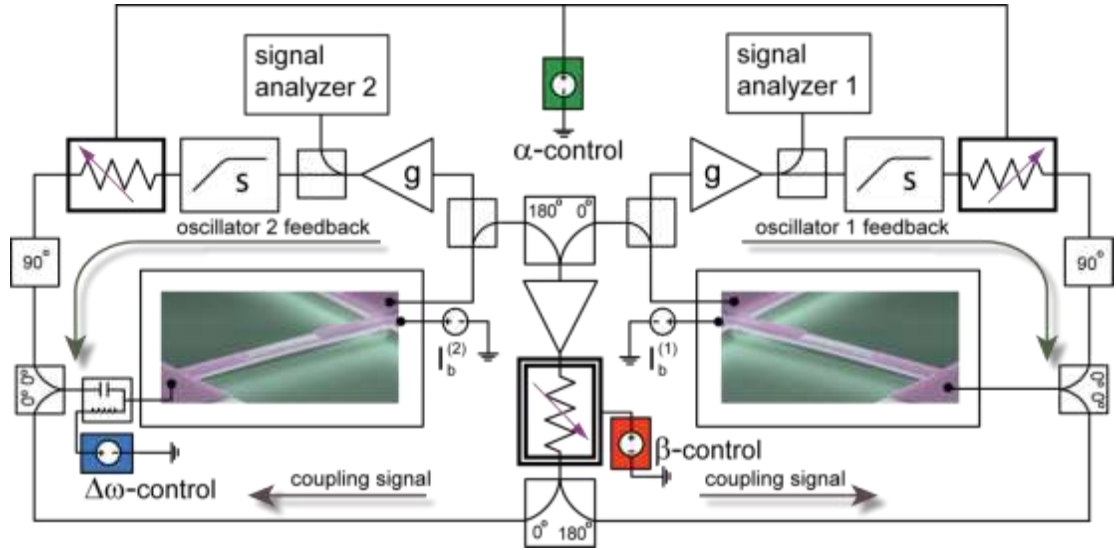

**Figure 1:** Simplified circuit schematic used for testing equations 1-3. The NEMS resonators employed were selected to be as identical as possible. Each NEMS resonator (colored SEM micrograph) is embedded in two feedback loops: one is used for creating self-sustained oscillations in each resonator, and the other for implementing coupling between the two oscillators. In the feedback loops, the signal is amplified with gain $g$ and then sent through a saturating limiter. The voltage controlled attenuators after each limiter (single heavy line boxes) in the feedback loop sets the level of oscillation, and constitutes a means to control the frequency pulling, $\alpha$, in equation 3, shown by the dc control in green. In the coupling loop the signal is amplified so that a voltage controlled attenuator (double heavy line boxes) adjusts the signal level in the common loop, thereby setting the coupling strength, shown by the dc control in red. The frequency difference is controlled by adjusting the stress induced in the left resonator by the piezovoltage shown in blue.

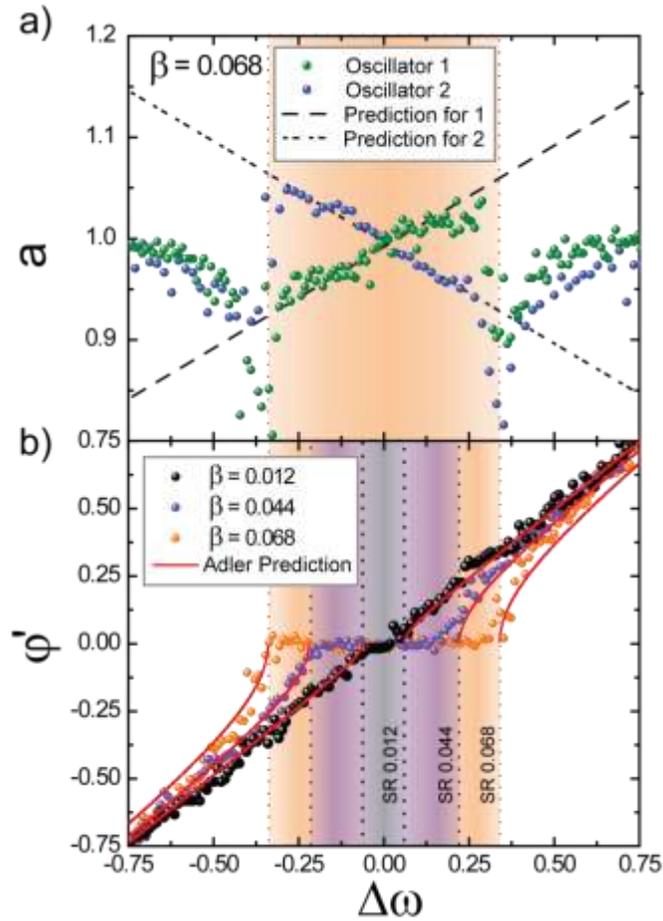

**Figure 2:** Synchronization in the limit of small coupling described by equations 5 and 6 with a frequency pulling $\alpha = 1.25$. **a)** Experimental data (points) are compared against theoretical predictions (lines) for the amplitudes of the two oscillators as the system moves through synchronization; the dependence upon detuning $\Delta\omega$ for a coupling of $\beta = 0.068$ is shown. The synchronized region is shown by orange shading. **b)** Data and predictions for the frequency difference $\varphi'$ for three different values of coupling. The set of data with the largest value of coupling $\beta = 0.068$ corresponds to the amplitude data from the upper plot. Frequency locking is clearly shown where values $\varphi' = 0$ occur. This defines the synchronization regime, which increases in size as the coupling increases. Predictions of equation 5 and 6are made by measuring the frequency pulling, coupling, and frequency detuning independently of the synchronization phenomenon. SR 0.012, SR 0.044, SR 0.068 denote the synchronized regimes (shaded regions) for the three couplings.

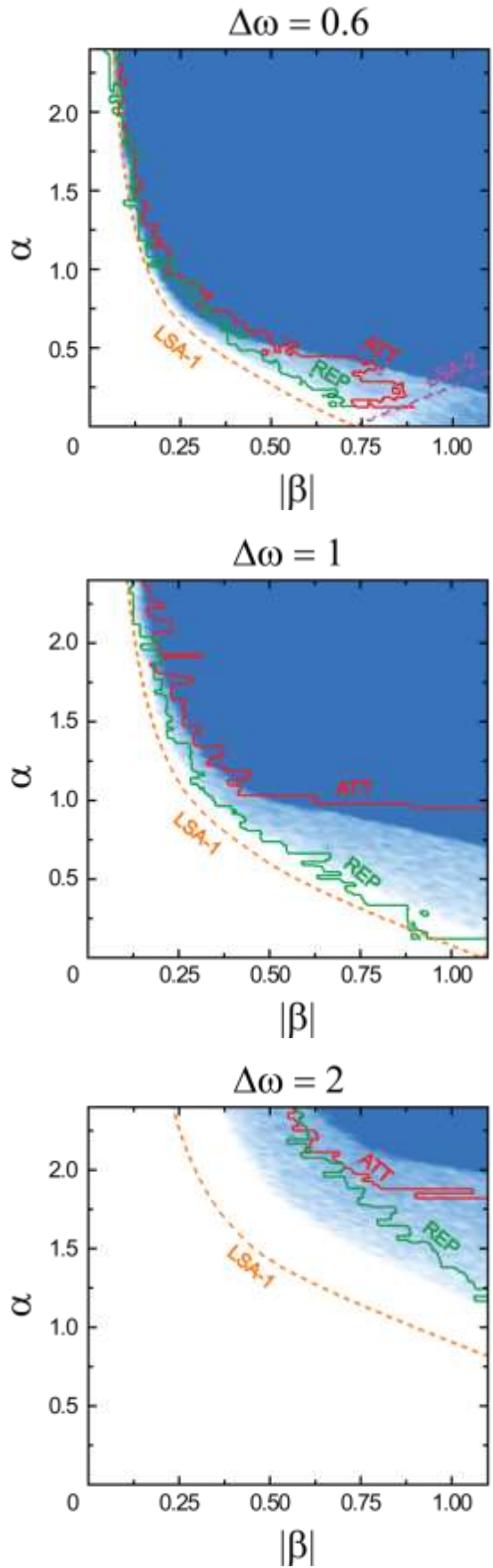

**Figure 3:** Experimentally measured synchronization space as a function of coupling $\beta$ and frequency pulling $\alpha$ for three different detunings $\Delta\omega = 0.6, 1, 2$. The basins of attraction, found from the time domain simulation of Equations 1-3, are shown by the gradient between white and blue, and correspond to the average number of times the simulation synchronized under 100 random initial phases. All lines show boundaries with respective regions to the right of the line. The green solid line (REP) is the experimental boundary for the transition from unsynchronized to the synchronized state for repulsive coupling. The red solid line (ATT) is the experimental boundary for the transition from the unsynchronized to the synchronized state for attractive coupling. The experimental synchronized state is defined as $\varphi' < 0.05$ (in units of the resonator width). The orange dashed line (LSA-1) depicts the theoretically-predicted (from a linear stability analysis) boundary for which at least one synchronized state (either in-phase or anti-phase, depending on coupling) is stable. Similarly, the purple dashed line (LSA-2) bounds the space for which both synchronized states are stable.

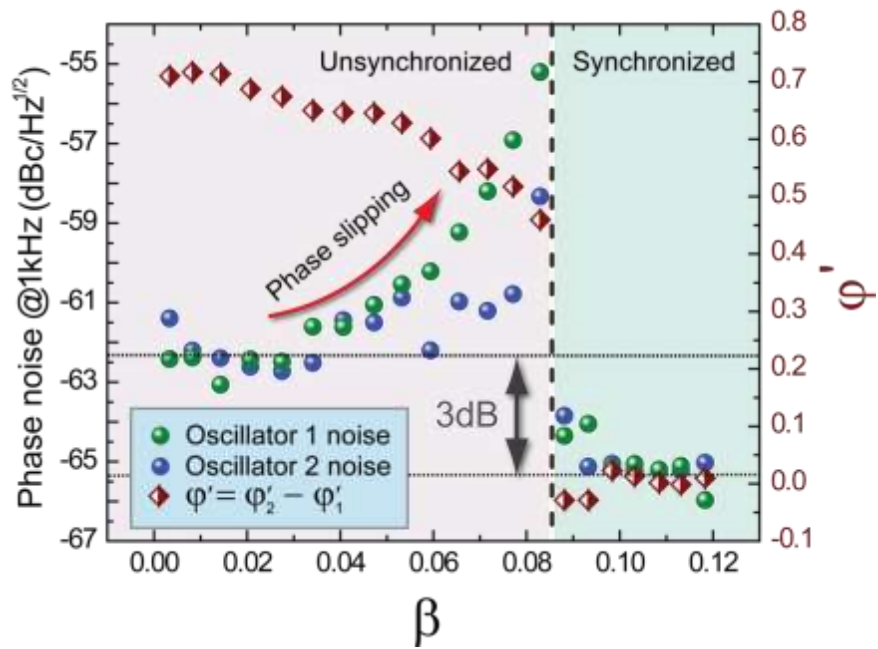

**Figure 4:** Oscillator phase noise at 1 kHz offset from carrier frequency (blue and green spheres, left axis) and oscillator frequency difference (red diamonds, right axis) as coupling is increased. At the value of coupling $\beta = 0.086$ the oscillator frequency difference goes to zero and the phase noise for both oscillators decreases by 3dB, i.e., corresponding to reducing the phase noise by half. This effect is due to noise averaging noted by Reference 1.

**Acknowledgements**

We thank E. Kenig and X.L. Feng for discussions and P. Ivaldi, E. Defaÿ and S. Hentz for providing us with the AlN/SOI material. M.C. Cross acknowledges financial support from the National Science Foundation grants DMR-0314069 and DMR-1003337.

*Supplementary Information*

# Synchronization of two anharmonic NEMS oscillators

M. H. Matheny, M. Grau, L. G. Villanueva, R. B. Karabalin, M. C. Cross, M. L. Roukes

*Kavli Nanoscience Institute and Departments of Physics, Applied Physics, and Bioengineering, California Institute of Technology, Pasadena, California 91125*

## I. *Theoretical derivation of synchronization equations for two anharmonic oscillators*

We start with the slow time equation[1] for two oscillators with two feedback[2, 3] terms: one which is common to both oscillators ($f_c$), and one that affects the corresponding oscillator only ($f_i$, i=1,2). These are

$$\frac{d\tilde{A}_1}{dT} - i\left(\frac{\delta_1}{2} + \lambda_{11}|\tilde{A}_1|^2\right)\tilde{A}_1 + \frac{\tilde{A}_1}{2} = f_1(\tilde{A}_1) + f_c(\tilde{A}_1, \tilde{A}_2), \quad \text{(S.I.1)}$$

and

$$\frac{d\tilde{A}_2}{dT} - i\left(\frac{\delta_2}{2} + \lambda_{22}|\tilde{A}_2|^2\right)\tilde{A}_2 + \frac{\tilde{A}_2}{2} = f_2(\tilde{A}_2) + f_c(\tilde{A}_2, \tilde{A}_1) \quad \text{(S.I.2)}$$

where $\delta_{1,2} = Q(\frac{\omega_{1,2}^2}{\omega_0^2} - 1)$ defines the offset of the natural resonance frequency of each NEMS device $\omega_{1,2}$ to a nearby frequency, $\omega_0$, $\tilde{A} = \tilde{a}e^{i\varphi}$ is the complex slow time oscillator displacement, and the nonlinear coefficient $\lambda$ is the relationship of the device displacement to the relative change in the NEMS resonance frequency (there is a factor of 3/8 that has been folded into this constant with respect to references 1 and 2). The terms $f_{1,2}$ and $f_c$ are the "oscillator feedback" and "coupling signal" in Figure 1, respectively. For a saturated oscillator feedback with linear, reactive, diffusive coupling these equations become

$$\frac{d\tilde{A}_1}{dT} - i\left(\frac{\delta_1}{2} + \lambda_{11}|\tilde{A}_1|^2\right)\tilde{A}_1 + \frac{\tilde{A}_1}{2} = \frac{s}{2}e^{i\varphi_1} + i\frac{\beta}{2}(\tilde{A}_2 - \tilde{A}_1), \qquad (\text{S.I.3})$$

and

$$\frac{d\tilde{A}_2}{dT} - i\left(\frac{\delta_2}{2} + \lambda_{22}|\tilde{A}_2|^2\right)\tilde{A}_2 + \frac{\tilde{A}_2}{2} = \frac{s}{2}e^{i\varphi_2} + i\frac{\beta}{2}(\tilde{A}_1 - \tilde{A}_2), \qquad (\text{S.I.4})$$

where $s$ is the level of the saturation, and $\beta$ is the (real-valued) strength of the coupling.

The magnitude of oscillation can be scaled by the saturation s ($\tilde{A} = A * s$), which yields

$$\frac{dA_1}{dT} - i\left(\frac{\delta_1}{2} + \lambda_{11}s^2|A_1|^2\right)A_1 + \frac{A_1}{2} = \frac{1}{2}e^{i\varphi_1} + i\frac{\beta}{2}(A_2 - A_1), \qquad (\text{S.I.5})$$

and

$$\frac{dA_2}{dT} - i\left(\frac{\delta_2}{2} + \lambda_{22}s^2|A_2|^2\right)A_2 + \frac{A_2}{2} = \frac{1}{2}e^{i\varphi_2} + i\frac{\beta}{2}(A_1 - A_2). \qquad (\text{S.I.6})$$

We combine the terms $\lambda_{11}s^2$ into a single term $\alpha$, which is the nonlinear frequency pulling[4].

Equations S.I.5 and S.I.6 can be separated into magnitude and phase,

$$\frac{da_1}{dT} = -\frac{a_1}{2} + \frac{1}{2} + \text{Re}\left(i\frac{\beta}{2}\left(a_2 e^{i(\varphi_2 - \varphi_1)} - a_1\right)\right), \qquad (\text{S.I.7})$$

$$\frac{da_2}{dT} = -\frac{a_2}{2} + \frac{1}{2} + \text{Re}\left(i\frac{\beta}{2}\left(a_1 e^{i(\varphi_1 - \varphi_2)} - a_2\right)\right), \qquad (\text{S.I.8})$$

$$\frac{d\varphi_1}{dT} = \frac{\delta_1}{2} + \alpha a_1^2 + \text{Im}\left(i\frac{\beta}{2}\left(\frac{a_2}{a_1}e^{i(\varphi_2 - \varphi_1)} - 1\right)\right), \qquad (\text{S.I.9})$$

and

$$\frac{d\varphi_2}{dT} = \frac{\delta_2}{2} + \alpha a_2^2 + \text{Im}\left(i\frac{\beta}{2}\left(\frac{a_1}{a_2}e^{i(\varphi_1-\varphi_2)} - 1\right)\right). \tag{S.I.10}$$

To examine synchronized states we look at the oscillator phase difference $\varphi = \varphi_2 - \varphi_1$. Equations S.I.7-S.I.10 become

$$\frac{da_1}{dT} = -\frac{a_1}{2} + \frac{1}{2} - \frac{\beta}{2}a_2 \sin\varphi, \tag{S.I.11}$$

$$\frac{da_2}{dT} = -\frac{a_2}{2} + \frac{1}{2} + \frac{\beta}{2}a_1 \sin\varphi, \tag{S.I.12}$$

and

$$\frac{d\varphi}{dT} = \frac{d\varphi_2}{dT} - \frac{d\varphi_1}{dT} = \frac{\delta_2}{2} - \frac{\delta_1}{2} + \alpha a_2^2 - \alpha a_1^2 + \frac{\beta}{2}\left(\frac{a_1}{a_2} - \frac{a_2}{a_1}\right)\cos\varphi. \tag{S.I.13}$$

These are equations 1,2, and 3 from the main text where the prime character ' represents the derivative with respect to slow time $T$, and $\frac{\delta_2}{2} - \frac{\delta_1}{2} = \Delta\omega$.

## II. *Experimental Methods*

Experimental Methods: Device fabrication has been previously described by Villanueva *et al*[2]. All measurements were taken at a pressure of less than 100mT through a balanced bridge technique (not pictured in the figure)[5] in order to reduce the effect of parasitic capacitances[3]. All three synchronization parameters are modified by external and independent DC voltage sources. The coupling strength can be controlled by adjusting the feedback gain in the coupling loop. We amplify and tune (using the red DC control voltage box in Figure 1 of main text) a voltage controlled attenuator (double-line box in Figure 1 of main text) in order to modify this coupling feedback gain. The frequency difference between the two oscillators can be linearly controlled by inducing stress in one of the beams by a DC piezovoltage, as shown by the blue box in Figure 1. The frequency pulling can be adjusted by varying the absolute oscillator amplitudes (keeping the relative amplitudes fixed to ensure that Equations 1-3 remain valid) through the use of the voltage controlled attenuators (single-line dark box controlled by green DC voltage). The amplifiers used in the setup were tested at each stage to ensure linearity of signal transfer. For more information see Supplementary Information section III. All data (except for phase noise) are taken by two separate spectrum analyzers so that amplitude and frequency can be measured independently for the two oscillators. Phase noise is measured on a single spectrum analyzer with a phase noise module, with the input switched for either oscillator. Simulations (Figure 3) of the basins of attraction were carried out in Matlab, and the linear stability analysis was carried out in Mathematica.

## III. Experimental resonator properties

The devices were selected such that the parameters were nearly identical. In Table S.1 we show the values for the resonator frequency and quality factor. We have published more details elsewhere[3]. Note that throughout the quality factors and frequencies varied ~6% from device heating due to piezoresistive bias.

| Parameter | Device "1" | Device "2" |
| --- | --- | --- |
| Frequency, $f_0$ | 13.056 MHz | 13.060 MHz |
| Q factor | 1640±70 | 1680±100 |

In Figure S.1 we show the driven response from which the values of Table S1 are found. The devices came from the same fabrication run and design.

Table S.1. Fit parameters of the resonance plots shown below. The uncertainties in Q are found from measuring Q at different times throughout the experiment.

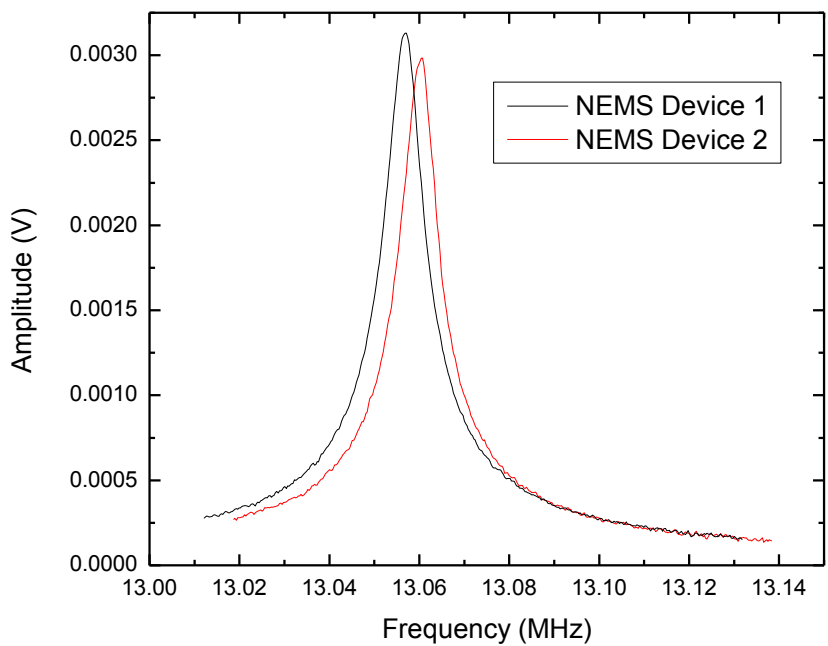

**Figure S.1:** Driven response of the two devices. Note the similarity in frequency and quality factor.

## IV. *Calibration of setup, and measurement of synchronization parameters*

Before calibrating the three synchronization parameters ($\Delta\omega, \alpha, \beta$), we must ensure that both the "oscillator" and "coupling" feedback signals (see Figure 1 of the main text) have the proper phase shifts, i.e., $f_{1,2}$ is purely dissipative and $f_c$ is purely reactive in equations S.I.1 and S.I.2. If the oscillators are uncoupled, the proper phase shift in the "oscillator" feedback loops causes maximum oscillation. At low saturation, the oscillator magnitude is a Lorentzian[3] function with respect to the frequency. In the slow time, this is

$$\left|\tilde{A}\right|^2 \propto \frac{s^2}{1+4\Omega^2}, \tag{S.II.1}$$

with $\Omega = \frac{d\varphi_{1,2}}{dT}$ from equations S.I.9 and S.I.10. We measure the oscillation amplitude and frequency as the phase shift in the oscillator loop is varied. We plot this for oscillator 1 in Figure S.2(a). A Lorentzian fit of this data yields the proper setting for the voltage controlled phase shifter embedded in the oscillator loop (the voltage controlled phase shifter is not pictured in Figure 1 from the main text). From the Lorentzian fit, the central frequency gives us the proper setting for the phase shifter, as shown in figure S.2(b).

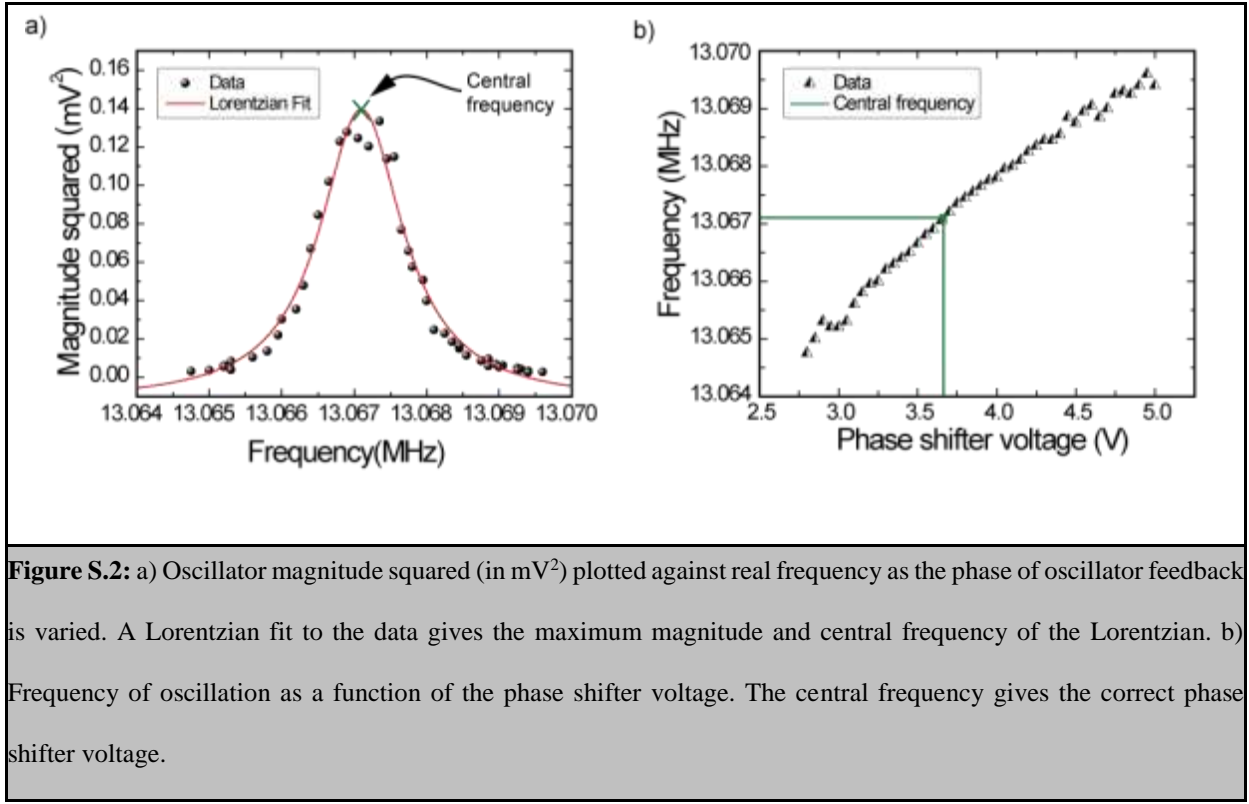

**Figure S.2:** a) Oscillator magnitude squared (in mV²) plotted against real frequency as the phase of oscillator feedback is varied. A Lorentzian fit to the data gives the maximum magnitude and central frequency of the Lorentzian. b) Frequency of oscillation as a function of the phase shifter voltage. The central frequency gives the correct phase shifter voltage.

### A. Calibration of coupling, β

In order to verify that the coupling loop is purely reactive, we compare two different measurements: 1) the level of amplification of the signal from the NEMS device through the coupling loop, and 2) the frequency shifts of the two oscillators due to the coupling feedback. Note that if the coupling is not strictly reactive, then according to reference 3, we must include a dissipative term to the feedback,

$$f_c(A_1, A_2) = (K + i\beta)(A_2 - A_1). \tag{S.II.2}$$

From figure 1 in the main text, if we turn off the second oscillator, then S.II.2 gives

$$f_c(A_1, A_2) = (K + i\beta)(-A_1). \tag{S.II.3}$$

The feedback described in equation S.II.3, when inserted into equation S.I.5, will lead to not only tuning of the oscillator by $-\beta/2$, but also additional dissipation proportional to $K$. Note that the magnitude $|K + i\beta|$ is the total gain of a signal through the coupling loop. By measuring this coupling loop gain and comparing it to measurements of the oscillator frequency shift, we can verify that $|K + i\beta| = |\beta|$, i.e., our coupling is strictly reactive. In Figure S.3, $|K + i\beta|$ is measured by the first method (red line, right vertical axis), and $\beta$ is measured using the frequency shift of the two oscillators (blue stars and green circles, left vertical axis). If the red line had a larger magnitude than the data points from the frequency shifts, then there would be a dissipative component to the feedback ( $K \neq 0$ ). However, these two measurements agree. These measurements not only verify that the coupling feedback has the proper phase shift, but also provide a calibration for coupling $\beta$ in terms of the voltage of the coupling attenuator.

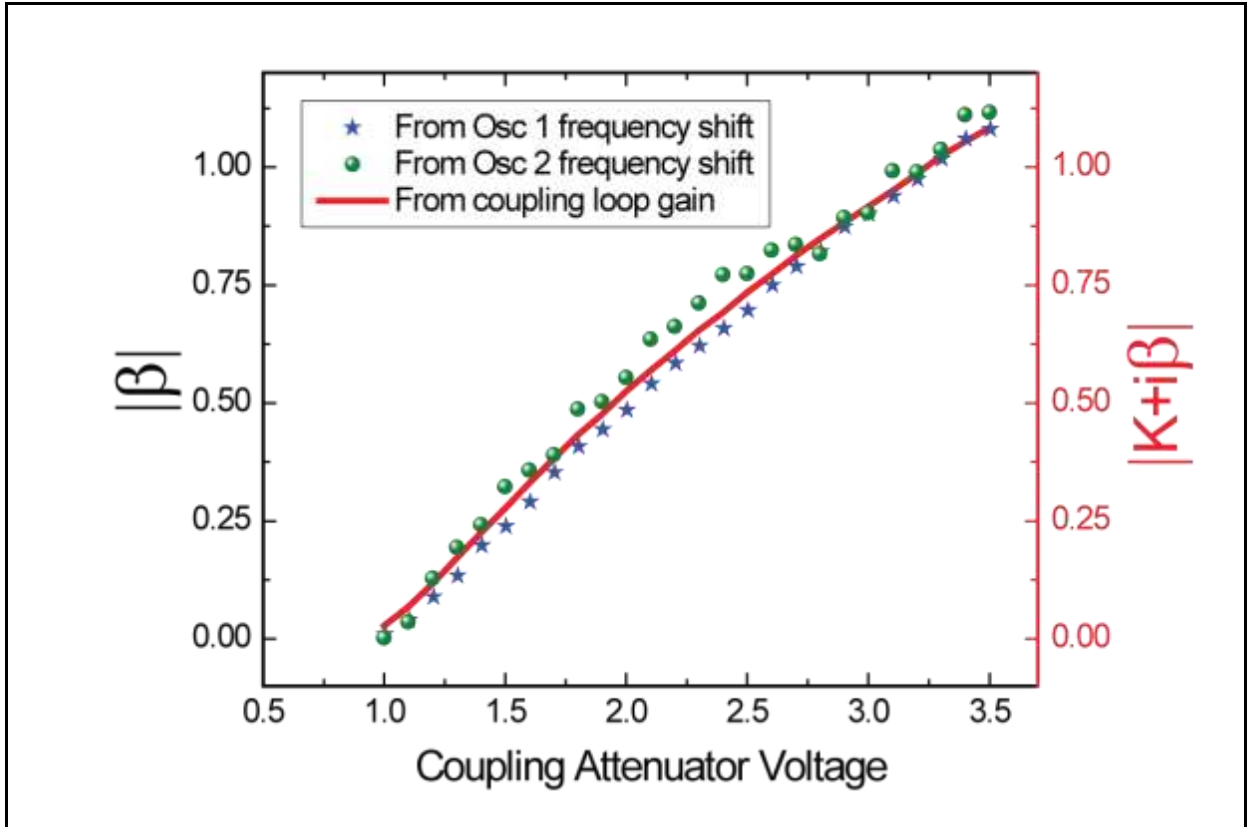

**Figure S.3.** Different measurements to calibrate coupling. The green (blue) points are found using tuning data from oscillator 1 (2), and correspond to the left vertical axis. The red curve is found by measuring gain around the coupling loop; it corresponds to the right vertical axis.

## B. Calibration of frequency pulling, $\alpha$

In order to calibrate the frequency pulling $\alpha = \lambda s^2$, we first calibrate the NEMS displacement and oscillator magnitude $|\tilde{A}|$. The thermomechanical noise of the NEMS device provides an absolute scale by which we can calibrate the device displacement from the electronic signal[2]. We can scale the NEMS displacement to the oscillator magnitude. With the oscillator and coupling feedback turned off, we measure the frequency response of the NEMS device under a constant level of external excitation. Fitting the NEMS frequency at the peak magnitude, for different values of excitation, yields the nonlinear coefficient $\lambda$[6]. hen the oscillators are uncoupled, the maximum oscillator amplitude corresponds to the level of saturation *s* (equation S.I.3 and S.I.4). Changes to the feedback saturation level, and thus the nonlinear pulling, can be made by adjusting the oscillator loop's attenuator after the limiting diode, as diagrammed in Figure 1 of the main text.

### C. Calibration of detuning, Δω

We present two different ways of measuring the detuning $\Delta\omega$. When detuning is held fixed, a low value of coupling $\beta$ in equations S.I.11-S.I.13 yields a phase equation

$$\frac{d\varphi}{dT} = \Delta\omega + \alpha a_2^2 - \alpha a_1^2 = \Delta\omega. \tag{S.II.4}$$

According to equation S.II.4, we can find the fixed detuning by measuring the oscillator frequency difference at zero coupling.

However, when the detuning is swept, a different calibration method is needed. In the experiment, we measure the oscillator frequency difference as a function of a piezoelectric tuning voltage (which changes the stress in one of the devices and hence the detuning[7]). We wish to make a correspondence between this tuning voltage and the detuning $\Delta\omega$. When the oscillators are far from the synchronization regime, the detuning dominates the other terms on the right hand side of equation S.I.13, so the oscillator frequency difference is proportional to the detuning. A linear fit to data far from the synchronized regime provides a relationship between the piezoelectric tuning voltage and the oscillator frequency difference. We therefore calibrate the detuning in terms of the piezoelectric tuning voltage by means of the oscillator frequency difference at data points far from synchronization. The calibration takes advantage of the fact that the detuning is linear in the piezoelectric tuning voltage[8]; we can interpolate each linear fit and calibrate the detuning in the synchronization regime. In figure S.4 (a), we show the raw data for the frequencies of the two oscillators from two sweeps with different coupling. In Figure S.4 (b), for the same sweeps, we show the (unscaled) oscillator frequency differences, where the linear fits are performed.

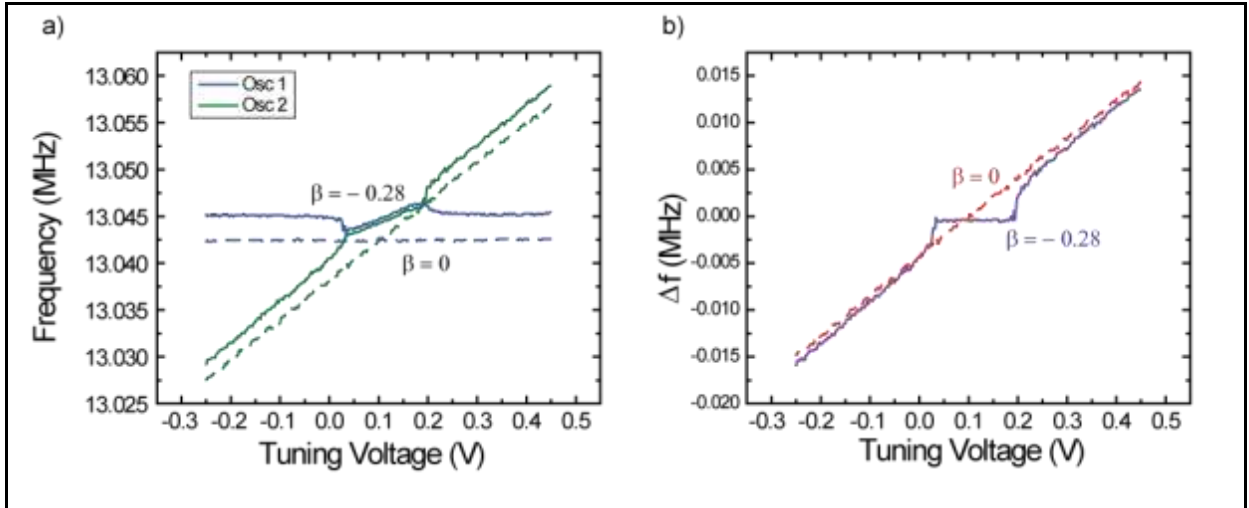

**Figure S.4:** a) Raw data of sweeps of detuning under different coupling conditions (dashed lines $\beta = 0$, solid lines $\beta = -0.28$), with each sweep taking minutes. The synchronization region appears when coupling is turned on. The two sweeps, taken hours apart, show that the NEMS device frequencies are drifting. b) Difference in frequency for the same sweeps. Straight lines are fit to the end sections of (b) in order to calibrate $\Delta\omega$ and correct for drifts.

The time between the measurements for the two different values of coupling ~hours, thus allowing drifts in oscillator frequencies to set in. However, the drift within each sweep is small, given that each sweep ~minutes. Therefore, through the method outlined above, each sweep can be calibrated to correct for these drifts.

Note that in Figure S.4 (a), with the coupling turned on, there is *mutual* entrainment, evidence that our coupling is symmetric. Adler's equation (equation 4 from the main text) originally described[9] an experiment where oscillator 1 is fed the signal of oscillator 2, but oscillator 2 is not fed the signal of oscillator 1. This asymmetric coupling led to oscillator 1, the "slave" oscillator, being dominated by oscillator 2, the "master" oscillator. In our experiment, it is clear that both oscillator frequencies shift towards one another, i.e., each oscillator has equal influence over the final state.

## V. *Notes on the phase noise measurement*

In the experiment, the oscillator phase noise is vastly different in the configuration shown for Figures 2 and 3 of the main text. When synchronization occurs in this setting, one oscillator dominates the noise of both when synchronized. However, if the phase delay of the oscillator feedback loops are adjusted, the phase noise of the two oscillators can be adjusted[3]. We adjust the feedback phase delay so that the phase noises are equivalent. Inevitably, a more general form of equations presented in Section I from the main text must be considered, and the values for alpha and delta omega cannot be calibrated as outlined in Section III. However, the coupling loops are not changed, and is very small and mutually symmetric, and so the overall behavior follows two simple phase oscillators. We therefore do not quantify when the synchronization will occur, but can predict the reduction in phase noise for the synchronized oscillators.

## VI. *Previous works on synchronization*

### A. *Definition of synchronization*

The widely accepted definition of synchronization is given in the text "Synchronization: A universal concept in nonlinear science" by Pikovsky, Rosenblum, and Kurths[10] on page 8 of the introduction:

"We understand synchronization as an **adjustment of rhythms of oscillating objects due to their weak interaction.**" *(emphasis theirs)*

They later expand on the concept of "weak interaction" on page 17:

"..we can say that the introduction of coupling should not qualitatively change the behavior of either one of the interacting systems and should not deprive the systems of their individuality."

And later on the same page:

"To call a phenomenon synchronization, we must be sure that:

- We analyse the behavior of two self-sustained oscillators, i.e. systems capable of generating their own rhythms;
- The systems adjust their rhythm due to a weak interaction;
- The adjustment of rhythms occurs in a certain range of systems' mismatch; in particular, if the frequency of one oscillator is slowly varied, the second follows this variation.

Correspondingly, a single observation is not sufficient to conclude synchronization. Synchronization is **a complex dynamical process, not a state.**" *(emphasis theirs).*

We examine the previous claims of synchronization with this definition.

## B. Previous claims of mechanical synchronization

We know of two prior claims of synchronization in miniaturized mechanical systems[11, 12]. These claims are examined in more detail in the following sections.

### a. Shim, et al. Science 2007

Shim, *et al.* claimed to observe the synchronization of a pair of coupled nanomechanical oscillators. However, that work studied a pair of coupled nanomechanical resonators driven by an external periodic signal and measured the response amplitude at the drive frequency or at a harmonic of the drive frequency. They did not give any experimental or theoretical evidence for self-sustained oscillations. The system which was under study had very strong coupling with the two linked beams always phase coherent. This study is analogous to a pair of pendulums with a rigid bar connecting the pendulum bobs, and driven with a harmonic force.

### b. Zhang, et al. PRL 2012

The configuration of the optomechanical system presented in Zhang, *et al.* is not capable of the weak coupling needed to establish synchronization. This is demonstrated both experimentally, and in the modeling. The authors also misinterpret a key piece of data. All figures mentioned in this section refer to the figures in Zhang, *et al.*

These authors performed their experiment by coupling two optomechanical oscillators together by their evanescent light fields. Before coupling them together, they confirmed their devices are individually oscillating, as shown Figures 3 a,b. Later, in Figures 3 c,d,e, they coupled the optical cavities together into symmetric and anti-symmetric normal modes, and excited the system through one device. This serves both to excite two oscillators simultaneously, with only one optical input and to couple the devices together.

The data in their Figure 3 show the system does not in fact have weak coupling. Figure 3 c,d,e shows the coupled system under three different values of laser power. Since an increase in the laser power is associated with an increase in the optical coupling between the two devices, Figure 3c is the data for the smallest coupling, and so if the system is strongly coupled in Figure 3c, the rest of the data is also strongly coupled. With respect to Figure 3c, the authors stated that the left optomechanical oscillator (L OMO) started self-sustained oscillation at the white dashed line. At a stronger laser detuning ( ~0.23 GHZ), the right optomechanical oscillator (R OMO) started self-sustained oscillation and the L oscillator shut off. This is also found in numerical simulation in Figure 3f. The fact that the oscillation of R OMO shut down L OMO is clear evidence the system was strongly coupled: the self-oscillation of one oscillator should not turn off the other, if they are truly independent and weakly coupled.

The model for the optomechanical system provided in the supplementary information of the study also demonstrates the strong coupling. In section S5.A the authors give equations for the radiation pressure induced self-oscillation, the optical spring effect, and the reactive and dissipative coupling between the oscillators. The model shows the dissipative coupling between oscillators is comparable to the driving terms of the self-oscillation, regardless of the laser power or detuning. Depending on the sign of the coupling, self-oscillation in one optomechanical device can turn off the other oscillator. This does not constitute weak coupling, since the behavior of the oscillators should not qualitatively change due to the interaction. This system, when the optical cavities are fully coupled into symmetric and anti-symmetric modes, is not capable of weak coupling. A separation of the coupling and the excitation mechanism is necessary for weak coupling a demonstration of synchronization.

The data in Figure 3d appears to show the two oscillations merging into a single oscillation; however, the authors misinterpreted the data in this figure, which does not in fact show evidence of two oscillations. The transition in Figure 3d is also presented in Figure 4 showing the spectra. The unsynchronized behavior in Figure 4d was suggested to show two independent self-sustained oscillations transitioning to a single synchronized state in Figure 4e. Examining the spectral width of the "R" oscillator (blue peak on the left in d) shows it is not consistent with self-sustained oscillation. As Zhang, *et al* pointed out earlier, when uncoupled oscillators are described, "…the optomechanical resonator starts self-sustaining oscillations and becomes an OMO **characterized by sudden linewidth narrowing** and oscillation amplitude growth." *(emphasis ours).* The spectral width of the blue peak in Figure 4d does not show this narrowing but is, however, consistent with the quoted width of the driven non-self-oscillatory resonance width, determined by the resonant frequency divided by the quality factor. In Figure 4d, the width of the blue peak on the left is approximately.

Also, the oscillations in Figure 3b are of a much different character (amplitude and width) than the one in Figure 3d for the "R" oscillator. It is also surprising, if the coupling was weak, that the two devices began self-oscillation at the same threshold in the coupled case (Figure 3d) when the thresholds were vastly different in the uncoupled case (Figures 3a,b): the authors do not give any explanation for this. We conclude that the publication by Zhang, *et al.* did not show two oscillations transitioning to a single oscillatory state.

### C. Josephson Junctions

There are two main features which distinguish our results from those of Josephson Junctions (JJs). Firstly, JJs behave as rotors driven by a constant torque, whereas our system is a more typical representation of standard self-sustained oscillators. Secondly, we have much more control over system parameters.

Josephson Junctions can be likened to a pendulum driven by a constant torque. The frequency of these rotations is a function of the applied torque exerted on the pendulum, which corresponds to an increase in the bias current across the JJ. However, in an oscillator, the frequency is determined by the physical properties of the system, such as pendulum length and gravitational restoring force for the pendulum. Although rotors share some features of self-sustained oscillators, there are important differences and they will not exhibit all the phenomenon found in self-sustained oscillations. Also, in JJs, the onset of the periodic motion is not a supercritical Hopf bifurcation as in simple feedback oscillators, but a saddle-node bifurcation[13].

The current state of the art for Josephson systems does not exhibit the degree of control as demonstrated in our system. Experiments on arrays of Josephson Junctions have demonstrated control over only the driving current, while we have control over all the parameters in our system. The JJ arrays can indeed be mapped to the Kuramoto model in some limit; however, extending their relevance to oscillator synchronization outside of this limit has not been shown. We have shown (Figure 3 of the main text) synchronization of self-sustained oscillations which do not obey the Kuramoto model. This phenomenon cannot be found in JJs.

## D. Spin-Torque Oscillators

Spin-torque oscillators have been shown to synchronize (see Kaka, *et al* 2005[14]), but the system control is limited and there is no match of any theoretical models to the experimental data, in contrast to our study. In Kaka, *et al.* only one parameter was changed, namely, the current, and therefore the frequency. The study had a fixed value of coupling (set by fabrication constraints). The presented implementation also has the disadvantage that changing the current (or the magnetic field) inevitably changes the power and noise properties. On the other hand in our system we can change the frequency independently of amplitude, noise, and coupling.